\documentclass[twocolumn,showpacs,preprintnumbers,amsmath,amssymb,superscriptaddress]{revtex4}

\usepackage{graphicx}
\usepackage{dcolumn}
\usepackage{bm}

\begin{document}

\preprint{APS/123-QED}

\title{
Coexistence of Bloch electrons and glassy electrons 
in Ca$_{10}$(Ir$_4$As$_8$)(Fe$_{2-x}$Ir$_x$As$_2$)$_5$  
revealed by angle-resolved photoemission spectroscopy
}

\author{K. Sawada}
\affiliation{Department of Complexity Science and Engineering and Department of Physics, 
University of Tokyo, 5-1-5 Kashiwanoha, Chiba 277-8561, Japan}

\author{D.~Ootsuki}
\affiliation{Department of Complexity Science and Engineering and Department of Physics, 
University of Tokyo, 5-1-5 Kashiwanoha, Chiba 277-8561, Japan}

\author{K.~Kudo}
\affiliation{Department of Physics, Okayama University, Kita-ku, Okayama 700-8530, Japan}

\author{D.~Mitsuoka}
\affiliation{Department of Physics, Okayama University, Kita-ku, Okayama 700-8530, Japan}

\author{M.~Nohara}
\affiliation{Department of Physics, Okayama University, Kita-ku, Okayama 700-8530, Japan}

\author{T.~Noda}
\affiliation{Department of Complexity Science and Engineering and Department of Physics, 
University of Tokyo, 5-1-5 Kashiwanoha, Chiba 277-8561, Japan}

\author{K.~Horiba}
\affiliation{Department of Physics, Okayama University, Kita-ku, Okayama 700-8530, Japan}

\author{M.~Kobayashi}
\affiliation{Institute of Materials Structure Science, High Energy Accelerator Research Organization (KEK), 
                 Tsukuba, Ibaraki 305-0801, Japan}

\author{K.~Ono}
\affiliation{Institute of Materials Structure Science, High Energy Accelerator Research Organization (KEK), 
                 Tsukuba, Ibaraki 305-0801, Japan}

\author{H.~Kumigashira}
\affiliation{Institute of Materials Structure Science, High Energy Accelerator Research Organization (KEK), 
                 Tsukuba, Ibaraki 305-0801, Japan}

\author{N.~L.~Saini}
\affiliation{Department of Physics, University of Roma "La Sapienza" Piazalle Aldo Moro 2, 00185 Roma, Italy}

\author{T.~Mizokawa}
\affiliation{Department of Complexity Science and Engineering and Department of Physics, 
University of Tokyo, 5-1-5 Kashiwanoha, Chiba 277-8561, Japan}
\affiliation{Department of Physics, University of Roma "La Sapienza" Piazalle Aldo Moro 2, 00185 Roma, Italy}

\date{\today}

\begin{abstract}
Angle-resolved photoemission spectroscopy of Ca$_{10}$(Ir$_4$As$_8$)(Fe$_{2-x}$Ir$_x$As$_2$)$_5$ 
shows that the Fe 3$d$ electrons in the FeAs layer form the hole-like Fermi pocket 
at the zone center and the electron-like Fermi pockets at the zone corners 
as commonly seen in various Fe-based superconductors.
The FeAs layer is heavily electron doped and has relatively good two dimensionality. 
On the other hand, the Ir 5$d$ electrons are metallic and glassy probably due
to atomic disorder related to the Ir 5$d$ orbital instability.
Ca$_{10}$(Ir$_4$As$_8$)(Fe$_{2-x}$Ir$_x$As$_2$)$_5$ exhibits a unique electronic state 
where the Bloch electrons in the FeAs layer coexist with the glassy electrons 
in the Ir$_4$As$_8$ layer.
\end{abstract}

\maketitle

\newpage


The discoveries of the superconductivity in the layered LaOFeP \cite{Kamihara2006} 
and LaOFeAs \cite{Kamihara2008} systems have induced intensive research activities 
on the physical properties of Fe pnictides. 
The Fe pnictide superconductors commonly have the FeAs layers, 
where each Fe ion is tetrahedrally coordinated by four pnictogen ions 
and the Fe ions form a square lattice.
The FeAs layers are separated by the spacer layers such as LaO, and modifications 
of the spacer layers often play essential roles in control of the superconducting 
properties of the FeAs layers.
Recently, Ca$_{10}$(Pt$_4$As$_8$)(Fe$_{2-x}$Pt$_x$As$_2$)$_5$
with Pt$_4$As$_8$ spacer layer has been discovered
\cite{Kakiya2011,Ni2011,Lohnert2011} and has been attracting great interest 
due to a possible interplay between the Pt$_4$As$_8$ layer and the FeAs layer.
Ca$_{10}$(Pt$_4$As$_8$)(Fe$_{2-x}$Pt$_x$As$_2$)$_5$ exhibits superconductivity
at 38 K and the Pt$_4$As$_8$ spacer layer is considered to provide electrons 
to the superconducting FeAs layer. The Pt$_4$As$_8$ spacer layer consists of
PtAs$_4$ square-planar geometry and is predicted to be metallic. 
Namely, the Pt 5$d$ orbitals hybridize with the Fe 3$d$ states 
at the Fermi level ($E_F$).
However, such hybridization between the Pt 5$d$ and Fe 3$d$ orbitals 
at $E_F$ is not clearly observed by angle-resolved photoemission
spectroscopy (ARPES), indicating that the Pt$_4$As$_8$ spacer layer 
is semiconducting. \cite{Neupane2012, Thirupathaiah2013} 
On the other hand, another recent ARPES study has shown that the Pt 5$d$ states 
have small contribution to the Fermi surfaces, partly consistent with the 
theoretical prediction. \cite{Shen2013}
In case of Ca$_{10}$(Pt$_4$As$_8$)(Fe$_{2-x}$Pt$_x$As$_2$)$_5$, the contribution 
of the Pt 5$d$ states at $E_F$ is rather small even though it may
exist as predicted by the theory.

Isostructural Ca$_{10}$(Ir$_4$As$_8$)(Fe$_{2-x}$Ir$_x$As$_2$)$_5$ is another member of 
the iron-based superconductors with $T_c$ = 16 K. \cite{Kudo2013}
This material has square-planer Ir$_4$As$_8$ layers, which are isotypic to 
the Pt$_4$As$_8$ layers in Ca$_{10}$(Pt$_4$As$_8$)(Fe$_{2-x}$Pt$_x$As$_2$)$_5$. 
The band-structure calculation for
Ca$_{10}$(Ir$_4$As$_8$)(Fe$_{2-x}$Ir$_x$As$_2$)$_5$ predicts that
the Ir 5$d$ orbitals have relatively large contribution to the density of states 
at $E_F$ compared to Ca$_{10}$(Pt$_4$As$_8$)(Fe$_{2-x}$Pt$_x$As$_2$)$_5$.
\cite{Kudo2013} In addition, a recent x-ray diffraction study by 
Sugawara {\it et al.} has reported doubling of the unit cell along the $c$-axis 
below 100 K which is attributed to the Ir 5$d$ orbital crossover. \cite{Sugawara2014} 
In this context, an ARPES study on Ca$_{10}$(Ir$_4$As$_8$)(Fe$_{2-x}$Ir$_x$As$_2$)$_5$
is highly interesting and important in order to reveal electronic structure
of the Ir$_4$As$_8$ layer as well as its impact on the FeAs layer.
In this work, on the basis of photoemission spectroscopy, 
we discuss the unusual electronic structure of 
Ca$_{10}$(Ir$_4$As$_8$)(Fe$_{2-x}$Ir$_x$As$_2$)$_5$
where Bloch-like Fe 3$d$ electrons forming the hole and electron Fermi pockets
coexist with metallic and glassy Ir 5$d$ electrons.

The single crystal samples of Ca$_{10}$(Ir$_4$As$_8$)(Fe$_{2-x}$Ir$_x$As$_2$)$_5$  
were prepared as reported in the literature. \cite{Kudo2013}
The lattice constants $a$ and $c$ are $\sim$ 8.725 $\AA$ and 20.70 $\AA$ 
around 20 K \cite{Sugawara2014}.
The ARPES measurements were performed at beamline 28A of Photon Factory, 
KEK using a SCIENTA SES-2002 electron analyzer with circularly polarized light. 
The total energy resolution was set to 20 - 30 meV for 
the excitation energies from $h\nu =$ 41 - 67 eV.
The base pressure of the spectrometer was in the $10^{-9}$ Pa range. 
The single crystals of Ca$_{10}$(Ir$_4$As$_8$)(Fe$_{2-x}$Ir$_x$As$_2$)$_5$,
oriented by {\it ex situ} Laue diffraction, were cleaved 
at 20 K under the ultrahigh vacuum and the spectra were acquired 
at 20 K within 12 hours after the cleaving. 
$E_F$ was determined using the Fermi edge of gold reference samples.

\begin{figure}
\includegraphics[width=0.3\textwidth]{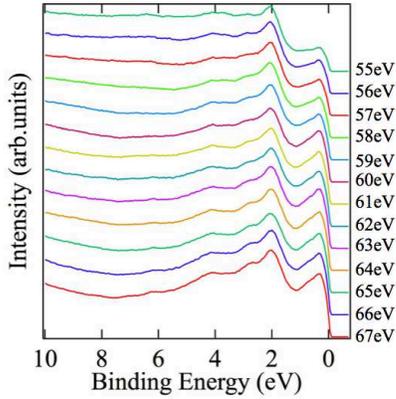}
\caption{(Color online)
Valence-band photoemission spectra of
Ca$_{10}$(Ir$_4$As$_8$)(Fe$_{2-x}$Ir$_x$As$_2$)$_5$  
taken at $h\nu =$ 55 - 67 eV.
}
\end{figure}

The valence-band photoemission spectra of Ca$_{10}$(Ir$_4$As$_8$)(Fe$_{2-x}$Ir$_x$As$_2$)$_5$
taken at photon energies from $h\nu$ = 55 eV to 67 eV are displayed in Fig. 1. 
The peak near $E_F$ dramatically gains its intensity in going from $h\nu$ = 55 eV to 67 eV, 
consistent with the Fe 3$p$-3$d$ resonance behavior observed in LaFeAsO$_{1-x}$F$_x$.
\cite{Malaeb2008} In contrast, the intensity of the region from 2 eV to 4 eV below $E_F$ 
does not depend appreciably on the photon energy. 
The valence-band spectra are basically consistent with the density of states calculated 
by Kudo {\it et al}. \cite{Kudo2013}
The calculation indicates that the structure near $E_F$ is dominated 
by the Fe 3$d$ orbitals with a mixture of the Ir 5$d$ $e_g$ orbitals,
consistent with the Fe 3$p$-3$d$ resonance.
On the other hand,
the structures ranging from 2 eV to 4 eV below $E_F$ can be assigned 
to the Ir 5$d$ $t_{2g}$ orbitals which are mixed with the Fe 3$d$ orbitals.
The overall agreement with the band-structure calculation suggests that
the Ir 5$d$ $e_g$ electrons contribute to the density states at $E_F$
as predicted by the calculation.

\begin{figure}
\includegraphics[width=0.48\textwidth]{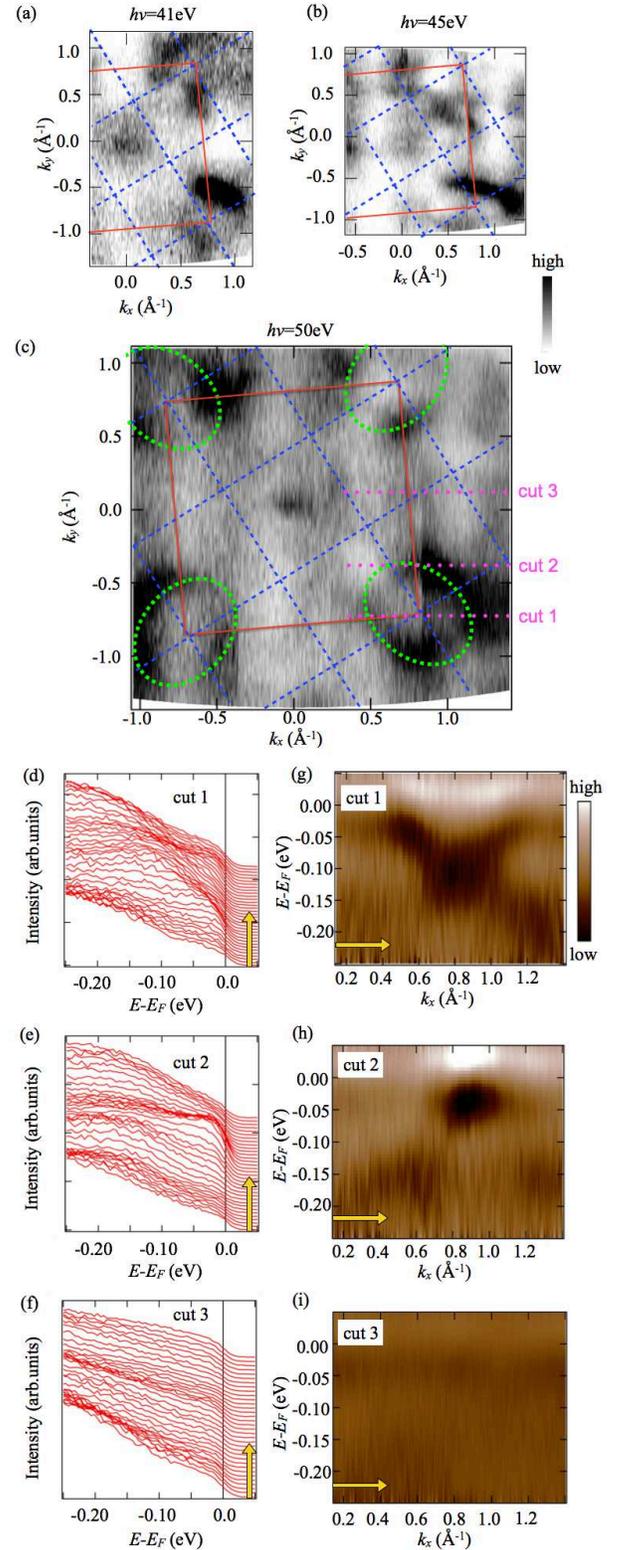}
\caption{(Color online)
Fermi surface maps (a) taken at $h\nu$ = 41.0 eV,
(b) taken at $h\nu$ = 45.0 eV, and (c) taken at $h\nu$ = 50.0 eV.
Two-dimensional Brillouin zones for the FeAs layer
and the Ir$_4$As$_8$ layer are indicated by the red solid
lines and by the blue dotted lines, respectively.
The electron-like Fermi pockets of the FeAs layer 
are indicated by the dotted ellipses around the zone corners.
Energy distribution curves (d) along cut 1, (e) along cut 2,
and (f) along cut 3 in the Fermi surface map at $h\nu$ = 50.0 eV.
Second derivative plots (g) along cut 1, (h) along cut 2,
and (i) along cut 3.}
\end{figure}

\begin{figure}
\includegraphics[width=0.48\textwidth]{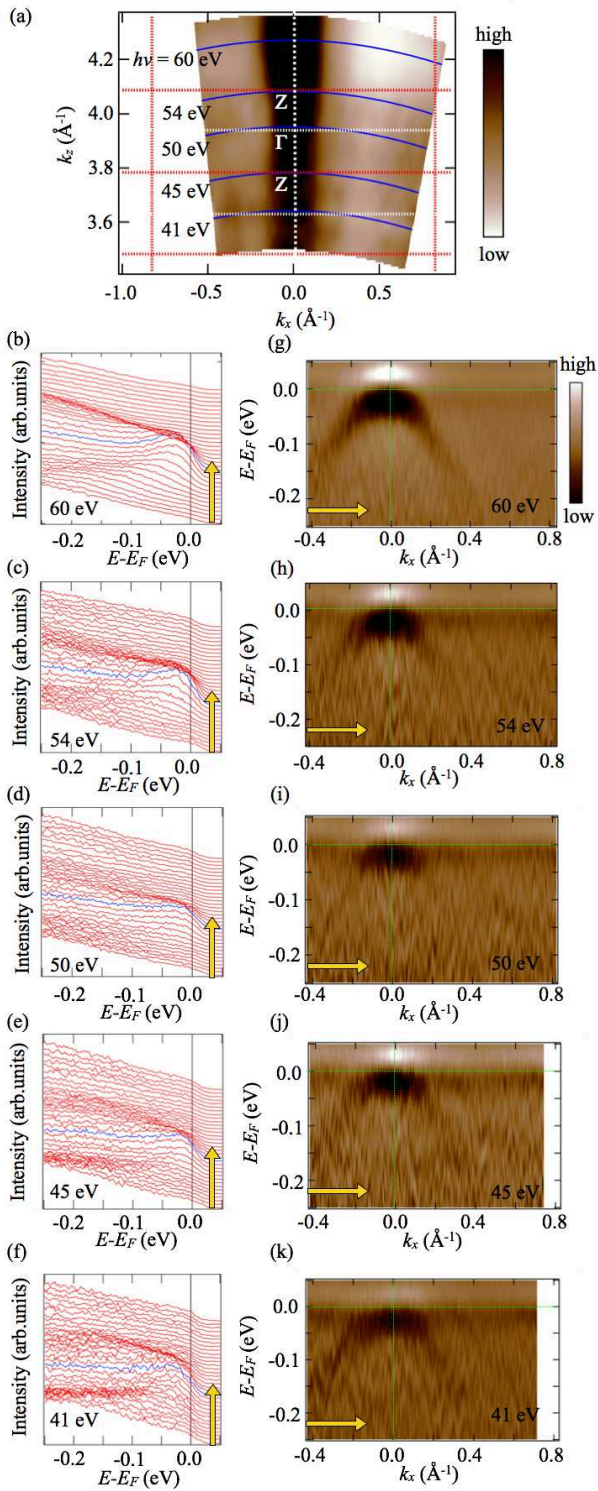}
\caption{(Color online) 
(a) Fermi surface map in the $k_x$-$k_z$ plane taken at $h\nu$ = 41 - 67 eV.
$k_x$ is the momentum approximately along the Fe-Fe direction of the FeAs layer.
$k_z$ is the momentum perpendicular to the FeAs layer.
Energy distribution curves (b) at $h\nu$ = 60 eV, (c) at $h\nu$ = 54 eV,
(d) at $h\nu$ = 50 eV, (e) at $h\nu$ = 45 eV, and (f) at $h\nu$ = 41 eV.
Second derivative plots (g) at $h\nu$ = 60 eV, (h) at $h\nu$ = 54 eV,
(i) at $h\nu$ = 50 eV, (j) at $h\nu$ = 45 eV, and (k) at $h\nu$ = 41 eV.
}
\end{figure}

Figures 2(a)-(c) show the Fermi surface maps taken at $h\nu$ = 41 eV, 
45 eV, and 50 eV. In the Fermi surface maps, the ARPES intensity $\rho(E)$ 
is integrated within energy window $\mid E-E_F \mid$ < 5 meV and is displayed 
as a function of $k_x$ and $k_y$. Here, $k_x$ (or $k_y$) is the electron momentum 
approximately along the Fe-Fe direction of the FeAs layer, and the two-dimensional 
Brillouin zone for the FeAs layer is indicated by the solid lines in Figs. 2(a)-(c).
The hole-like Fermi pocket at the zone center and the electron-like Fermi 
pockets at the zone corners are observed by ARPES as commonly seen 
in various Fe-based superconductors.
Here, one can safely conclude that the Fermi pockets are derived from the FeAs layer.
The Fermi surface maps of Figs. 2(a), (b), and (c) are taken at
$h\nu$ = 41 eV, 45 eV, and 50 eV, which correspond to $k_z$ 
(momentum perpendicular to the FeAs layer or the Ir$_4$As$_8$ layer) 
of 3.64 $\AA^{-1}$ ($\sim$ $12\times2\pi/c$), 3.78 $\AA^{-1}$ ($\sim$ $12.5\times2\pi/c$),
and 3.95 $\AA^{-1}$ ($\sim$ $13\times2\pi/c$) at the zone center ($k_x = k_y =0$), respectively. 
Whereas the area of the hole-like Fermi pocket at the zone center is very small,
those of the electron-like Fermi pockets at the zone corners are relatively large
for all the $k_z$ values, indicating that the FeAs layer is heavily electron doped. 
On the other hand, the Fermi surfaces from the Ir$_4$As$_8$ layer are 
apparently absent which may contradict with the theoretical prediction of 
the metallic Ir$_4$As$_8$ layer.

Figures 2(d), (e), and (f) show the ARPES intensity or 
the energy distribution curves $\rho(E)$  along cuts 1, 2, and 3 
in the Fermi surface map of Fig. 2(c), respectively. 
In Figs. 2(g), (h), and (i), the second derivatives $d^2\rho(E)/dE^2$
are plotted as functions of the momentum along cuts 1, 2, an 3.
The low value of $d^2\rho(E)/dE^2$ corresponds to the peak in $\rho(E)$.
Firstly, dispersion of the electron-like band is clearly seen in Figs. 2(d)
as well as in the second derivative plot in Fig. 2(g).
The electron-like Fermi pocket is still observed along cut 2 showing that
the area of the electron-like Fermi pockets are relatively large as
schematically indicated by dotted ellipse in Fig. 2(c).
The bottom of the electron band at the zone corner is located around 0.1 eV
below $E_F$ which is very far from $E_F$ compared to 
those of the various Fe-based superconductors. For example, the bottom of 
the electron band at zone corner is located 
around 0.03 eV below $E_F$ in optimally electron doped 
BaFe$_{1-x}$Co$_x$As$_2$ \cite{Sudayama2011}. 
This observation is again consistent with the heavy electron doping in 
the FeAs layer. 
Secondly, even outside of the electron-like Fermi pockets,
substantial spectral weight at $E_F$ is observed.
As shown in Fig. 2(f), the spectral weight at $E_F$ is seen 
even along cut 3 where no Fermi surface is expected and is observed indeed 
[also see the second derivative plot in Fig. 2(i)].
The momentum independent spectral weight at $E_F$
indicates glassy (and metallic) electronic states which may 
be induced by atomic disorder. Since the electron-like and hole-like
Fermi pockets are derived from the FeAs layer, 
the glassy spectral weight can be attributed to the Ir$_4$As$_8$ layer
which may be atomically disordered by the Ir 5$d$ $e_g$ orbital instability.
This picture is indeed consistent with the extended x-ray absorption fine
structure at the Ir $L$ edge \cite{Saini2014}.

Figure 3(a) shows a Fermi surface map in the $k_x$-$k_z$ plane.
The hole-like Fermi pocket around the zone center is clearly
seen in the wide $k_z$ region, and the $k_z$ dependence of the
Fermi pocket is rather small. This indicates that two dimensionality
of the Fe 3$d$ bands is relatively good in the present system
and would be inconsistent with the metallic Ir$_4$As$_8$ layer 
since the metallic spacer layer is expected to provide strong interlayer 
interaction.
However, if the metallic Ir$_4$As$_8$ layer is strongly disordered
as discussed in the previous paragraph,
the interlayer interaction becomes momentum independent and, consequently,
the FeAs layer can exhibit good two dimensionality although the Ir$_4$As$_8$ 
interlayer is metallic.

Figures 3(b)-(f) show energy distribution curves taken at
$h\nu =$ 60 eV, 54 eV, 50 eV, 45 eV, and 41 eV.
In addition to the hole-like Fermi pocket at the zone center,
the momentum independent spectral weight is seen at $E_F$.
This is again consistent with the picture that the momentum dependent 
Fermi pocket is derived from the FeAs layer and that the momentum independent 
states are assigned to the atomically disordered Ir$_4$As$_8$ layer.
The band dispersion of the hole band is seen in the second derivative plots
of Figs. 3(g)-(k).
Only one hole band is observed although three hole bands (one Fe 3$d$ $xy$
band and two Fe 3$d$ $yz/zx$ bands) are expected from the theoretical 
calculations. At the zone center, the As 4$p_z$ orbitals of the FeAs layer
hybridize with the Fe 3$d$ $xy$ orbital and the Ir 5$d$ $3z^2-r^2$ orbital
in the neighboring Ir$_4$As$_8$ layer.
Here, one can speculate that the Fe 3$d$ $xy$ orbital is strongly
affected by the orbital crossover between $3z^2-r^2$ and $xy$ 
in the Ir$_4$As$_8$ layer \cite{Sugawara2014} and is substantially 
broadened due to the Ir atomic (and orbital) disorder.
On the other hand, the Fe 3$d$ $yz/zx$ orbitals are less affected 
by the Ir$_4$As$_8$ layer and are observed by ARPES. 
This assignment is further supported by the fact that the dispersion
of the observed hole band agrees well with that of the Fe 3$d$ $yz/zx$
hole band reported in various systems.

In conclusion, we have studied the electronic structure of
Ca$_{10}$(Ir$_4$As$_8$)(Fe$_{2-x}$Ir$_x$As$_2$)$_5$ in which
the metallic Ir$_4$As$_8$ layer with orbital degrees of freedom
is expected to play important roles. The ARPES results indicate
that the Ir 5$d$ electrons are metallic and glassy probably due
to atomic disorder related to the Ir 5$d$ orbital instability
\cite{Sugawara2014}.
On the other hand, the Fe 3$d$ electrons in the FeAs layer 
form the hole-like Fermi pocket at the zone center and 
the electron-like Fermi pockets at the zone corners 
as commonly seen in various Fe-based superconductors.
While the hole band with the Fe 3$d$ $yz/zx$ character is clearly observed,
other hole bands are smeared out probably due to the strong out-of-plane
disorder from the Ir$_4$As$_8$ layer.
The ARPES results indicate that the FeAs layer is heavily electron doped 
and has good two dimensionality. 
Ca$_{10}$(Ir$_4$As$_8$)(Fe$_{2-x}$Ir$_x$As$_2$)$_5$ is characterized 
by the coexistence of the Bloch electrons in the FeAs layer and 
the glassy electrons in the Ir$_4$As$_8$ layer at $E_F$.

The authors would like to thank the valuable discussion with Dr. N. Katayama 
and Prof. H. Sawa.
This work was partially supported by Grants-in-Aid from the Japan Society of 
the Promotion of Science (JSPS) (No. 22540363, No. 25400372, No. 24740238, and No. 25400356)
and the Funding Program for World-Leading Innovative R\&D on Science and Technology (FIRST Program) from JSPS.
The synchrotron radiation experiment was performed with the approval of 
Photon Factory, KEK (2013G021).

\end{document}